\documentstyle[aps]{revtex}

\begin{document}

 \title{BOSE-EINSTEIN CONDENSATION AND SUPERFLUIDITY IN TRAPPED ATOMIC GASES}

\author{Sandro  Stringari}
\address{Dipartimento di Fisica, Universit\`{a} di Trento,}
\address{and Istituto Nazionale per la Fisica della Materia,}
\address{I-38050 Povo, Italy}
  
\date{\today}
 
\maketitle
 
\begin{abstract}
\noindent

Bose-Einstein condensates confined in traps exhibit unique features which have
 been the object of 
extensive experimental and theoretical studies in the last few years. In this 
paper I will discuss
some issues concerning the behaviour of the order parameter and  the 
dynamic and superfluid effects exhibited by such systems. 

\end{abstract}


\subsection{Introduction}

The aim of these lectures is to review some 
 key features exhibited by Bose-Einstein condensation in 
trapped atomic gases, with special emphasis to  the phenomena of superfluidity 
which have been the 
object of recent experimental and theoretical investigation
and which are characterized in a non trivial way
 by the
combined  role of the phase of the order parameter and of two-body interactions. 

Let us start recalling that these systems are characterized, at sufficiently low
temperatures, by  a complex order parameter which can be written as

\begin{equation}
\Psi_0({\bf r},t) = \sqrt{n_0({\bf r},t)}e^{iS({\bf r},t)}
\label{orderpar}
\end{equation}
where $n_0=\mid \Psi_0\mid^2$ is the so called condensate density and $S$ is the phase. 
In general the condensate density does not coincide with 
the density $n$ of the gas, because of  the occurrence of    
quantum and thermal  effects. However, in a very dilute gas, 
where the quantum depletion is negligible and for  low temperatures, 
such that the thermal depletion 
is small, 
the condensate density  coincides in practice with the total density. For this reason one often identifies
$n_0$ with $n$. For the same reason in these systems the condensate density can be measured with
good precision through imaging techniques.  Even at higher 
temperature the condensate density can be extracted  experimentally 
with reasonable accuracy. 
In fact, due to the harmonic shape of the external confinement, the thermal 
component is spacially  separated from the condensate so that the density profile is characterized by a 
typical bimodal 
structure allowing for systematic experimental analysis. 
The order parameter of trapped atomic gases then emerges as a crucial quantity not only from the 
conceptual point of view like in all many-body systems characterized by a
broken symmetry, but also 
from the experimental one. This represents a key difference with respect to superfluid  
helium where the order parameter is  a very hidden variable which can be
measured  only under  very  special conditions. 
The order parameter of a trapped gas can vary in space, time and temperature and characterizes in a 
direct way most of the physical quantities that one measures in these systems. 

In the first part I will mainly discuss  the properties of the modulus of 
the order parameter.  From the theoretical point of view this 
quantity is interesting because even its 
shape at equilibrium is very sensitive to the presence of two-body interactions. In 
the second part 
I will give more emphasis on  the role of the phase, a 
quantity which plays a crucial role in characterizing the coherence and 
superfluid features of  our 
systems. Of course the modulus and  the phase of the order parameter 
are deeply related quantities and 
in many cases their behaviour cannot be discussed in a separate way.
Many of the results discussed in these notes are derived in more detail in ref.\cite{RMP} to which
we refer also for a more complete list of references.

The discussion on Bose-Einstein condensation   presented 
in these notes is stimulated by several features of fundamental interest that are
worth mentioning. In particular  it is important to recall that
  
-  BEC gases in traps 
provide a unique opportunity to study the  transition from the {\bf microscopic} to 
the {\bf macroscpic} world.

- In these systems BEC shows up both in {\bf momentum} and {\bf coordinate} space.

- BEC gases provide an almost ideal realization of  {\bf classical matter waves},
 giving rise to new coherence phenomena.

- Trapped gases are well suited to study {\bf rotational superfluid} phenomena.

\subsection{BEC is a phase transition}

In the presence of  harmonic trapping
\begin{equation}
V_{ext} ={1\over 2}m\left(\omega^2_xx^2 + \omega^2_yy^2 + \omega^2_zz^2\right)
\label{Vext}
\end{equation}
the critical temperature for Bose-Einstein condensation 
of an ideal gas can be easily evaluated using the standard techniques of 
quantum statistical mechanics. To this purpose one has to define
the thermodynamic limit. In the case of harmonic trapping the natural choice is to take
$N\to \infty$ and $\omega_{ho}\to 0$, with the combination $N\omega_{ho}^3$ kept fixed.
Here $\omega_{ho} =(\omega_x\omega_y\omega_z)^{1/3}$ is the geometrical 
average of the three harmonic frequencies. Notice that in this limit the system, differently from traditional
 thermodynamic bodies,
 is characterized by a non uniform density varying on a macroscopic
scale, much larger than the typical interatomic distance. 
The critical temperature takes the value 
\begin{equation}
k_BT_c = 0.94 \hbar \omega_{ho}N^{1/3}
\label{Tc}
\end{equation}
while the condensate fraction obeys the law
\begin{equation}
{N_0\over N} = 1-\left({T\over T_c}\right)^3 
\label{N0}
\end{equation}
for $T<T_c$.
Similarly to the case of the uniform gas one predicts a phase transition even in the absence of  two-body 
interactions. This unique prediction of quantum statistical  
mechanics can be now tested 
experimentally in  trapped Bose gases \cite{Ensher,Mewes}. 
In fact, close to the critical temperature, these gases are so 
dilute that one can neglect in first  approximation the role of
the interaction. Estimate (\ref{Tc})
has actually guided experimentalists to the right domain of temperatures  which turns out
to be of the order of $\mu K$ (typically the trapping 
frequencies are  of the order of a few $nK$ while the number $N$
 of atoms varies from $10^4$ to $10^7$). Also the temperature 
dependence of the condensate fraction turns out to be in reasonable agreement with the theoretical 
prediction (\ref{N0}) as shown by fig 1. 

An interesting question  is how two-body forces modify the ideal gas estimate 
(\ref{Tc}).  Interaction effects  can have two different origins. A first one is a 
genuine many-body effect exhibited by uniform systems where one works at fixed density. 
This 
effect is highly non trivial from the theoretical point 
of view \cite{Baym} but the resulting correction is  expected to be   small and  
difficult to observe in 
trapped gases. A second effect is the result of mean field forces which, in the case 
of positive scattering lengths, tend to push the gas towards the external region, thereby reducing the density of the 
gas. This causes a natural decrease of the critical temperature. The opposite happens
 if the interaction is attractive.
The mean field effect can be easily  calculated 
and the result takes the form \cite{Giorgini}
\begin{equation}
{\delta T_c \over T_c } = -1.3 {a\over a_{ho}}N^{1/6}
\label{deltaTc}
\end{equation}
where $a$ is the scattering length and $a_{ho}=\sqrt{\hbar/m\omega_{ho}}$
is the oscillator length.
Typical values for $\delta T_c/T_c$ are a few percent in the available traps and the present accuracy of 
experimental data does not allow for a quantitative check of this prediction. By tuning the value of
the scattering length to larger values it should  be possible 
to obtain more sizable corrections
to the value of the  critical temperature. 

The above discussion has introduced a first important combination of the 
relevant parameters of the problem, given by $N^{1/6}a/a_{ho}$. This
combination  fixes 
the importance of interaction effects on the thermodynamic  properties of 
the system.
Its origin is easily understood if one calculates the 
ratio between  the interaction energy  $E_{int} \sim gn$  
where $g=4\pi \hbar^2a/m$ is 
the  coupling constant fixed by the scattering length and $n$ is the 
density of the thermal cloud evaluated in the center 
of the trap,
and the thermal energy $E_T \sim k_BT$. A simple estimate for the thermal density 
is given by the classical value
 $n \sim N (k_BT /m\omega_{ho})^{3/2}$.
Using result (\ref{Tc}) 
for the critical temperature, one finds that the ratio 
$E_{int}/E_T$, evaluated at $T \sim T_c$, is actually fixed by the combination 
 $N^{1/6}a/a_{ho}$. 
 This should not be confused with 
the so called Thomas-Fermi combination $N a/a_{ho}$ which instead 
characterizes the effects of two body-interactions on the  zero 
temperature properties of a trapped condensate.
The Thomas-Fermi  combination 
is recovered by evaluating 
 the ratio between the interaction energy $E_{int} \sim gn$, where $n$ 
is now the  
central value of the condensate density, and the quantum oscillator energy 
$\hbar\omega_0$. 
 The dependence on $N$ given by the two combinations discussed 
 above is extremely  different and explains why the effects of the 
 interaction at $T \sim T_c$  are small, while at $T\sim 0$ they
 are so large that  cannot be treated in a perturbative way. 
 The role of the Thomas-Fermi parameter will 
 be discussed systematically in the next sections.

Before concluding this section it is useful to remind that the 
measurements of the temperature dependence of the condensate density 
available in trapped Bose gases are  more reliable  
than the ones obtained in superfluid helium \cite{Sokol}
(see fig.2). In helium the condensate 
fraction is not a natural observable and can be extracted only through 
elaborated analysis of neutron scattering data at high momentum 
transfer. Actually the main evidence for the phase transition
 in superfluid helium comes from the analysis of other thermodynamic
 quantities, like  the viscosity 
 and the specific heat. 

\subsection{Order parameter and  long range order}

A key quantity characterizing the  long range order of a many-body system 
is the one-body density matrix \cite{Penrose}:
\begin{equation}
n({\bf r},{\bf r}^{\prime}) = \langle \hat{\Psi}^{\dagger}({\bf r})\hat{\Psi}
({\bf r}^{\prime})\rangle
\label{n1}
\end{equation}
where $\hat{\Psi}({\bf r}) = \hat{a}_0\phi_0({\bf r}) + \sum_{i\ne 0} 
\hat{a}_i\phi_i({\bf r})$
is the field operator expressed in terms of the annihilation operators
relative to  a generic basis of single 
particle wave functions. We have here 
separated the contribution arising from the 
lowest single particle state $\phi_0$  
in order to emphasize the effect of BEC. 

Let us first  consider an ideal gas at zero temperature. In this 
case all the particles occupy the same state $\phi_0$ determined by the
solution of the Schr\"odinger equation for the single particle Hamiltonian
$-\hbar^2 \nabla^2/2m + V_{ext}$. The 
density matrix (\ref{n1}) then  takes the separable form
$n^{(1)} ({\bf r},{\bf r}^{\prime}) =
N\phi_0^*({\bf r})\phi_0({\bf r}^{\prime})$ and 
remains different from zero for macroscopic distances 
$\mid {\bf r}-{\bf r}^{\prime}\mid$ of the order 
of the size of the sample (long range order). This 
is also called first order coherence and is a key consequence of
Bose statistics. 

At finite temperatures one should include the thermal occupation  
of the other single particle states and eq.(\ref{n1}) yields
\begin{equation}
n({\bf r},{\bf r}^{\prime})  = N_0\phi_0^*({\bf r})\phi_0({\bf r}^{\prime}) 
+
\sum_{i\ne 0}n_i\phi_i^*({\bf r})\phi_i({\bf r}^{\prime})
\label{n1T}
\end{equation}
where $n_i =(\exp\beta(\epsilon_i-\mu)-1)^{-1}$ are the thermal occupation
 numbers and $\epsilon_i$ are the single-particle excitation energies 
 of the excited states $\phi_i$. In the thermodynamic
limit the ratio $N_0/N$ tends to a constant value (condensate fraction) giving rise to long range order, while
the  second term in the r.h.s. of eq.(\ref{n1T}) can be replaced with an integral and 
consequently  vanishes at large distances. The importance of long range order is hence reduced at
finite temperature 
since $N_0$ is smaller than at $T=0$. 
For a uniform ideal gas  
the decay of the thermal component is given by
$n(s) \to kT m /(4\pi s)$ where $s=\mid{\bf r}-{\bf r}^{\prime}\mid$. 
Above the critical temperature
one has no more long range order ($N_0/N \to 0$) and the one-body density matrix decays faster
than $1/s$ at large distances.
In the  limit $T\gg T_c$ one finds 
the gaussian  
behaviour $n(s) = n(0)\exp[-s^2mk_BT/2\hbar^2]$.

The effects due to Bose-Einstein condensation discussed above 
 can be  also derived  by  setting 
$\hat{a}_0 = \hat{a}^{\dagger}_0 \equiv \sqrt{N_0}$. This is the 
well known Bogoliubov prescription which corresponds 
to ignoring the non commutativity  
 between the  particle 
operators $a_0$ and $a^{\dagger}_0$ and treating them 
like $c$-numbers. In virtue of the Bogoliubov assumption 
the field operator takes the form 
\begin{equation}
\hat{\Psi}({\bf r}) = \Psi_0({\bf r}) +  \sum_{i \ne 0}
 \hat{a}_i\phi_i({\bf r})
\label{PsiB}
\end{equation}
where
\begin{equation}
\Psi_0({\bf r}) = \sqrt{N_0}\phi_0({\bf r})
\label{OP}
\end{equation}
is a classical field, called the order parameter.  The order parameter 
can be also regarded as the expectation value of the field operator
\begin{equation}
\Psi_0({\bf r}) = \langle \hat{\Psi}({\bf r})\rangle
\label{OP2}
\end{equation}
where the average is taken on a configuration with broken 
gauge symmetry. 
In terms of the order parameter the diagonal term of the density matrix  
can be written
as
\begin{equation}
n({\bf r}) = n_0({\bf r})+ n_T({\bf r})
\label{n0T}
\end{equation} 
where $n_0({\bf r}) = \mid\Psi_0({\bf r})\mid^2$ is the condensate density
and $n_T({\bf r}) = \sum_{i\ne 0}n_i\mid \phi_i({\bf r})\mid^2$ is
 the contribution arising from the particles out of the condensate. In a dilute gas 
$n_T$ can be safely identified with 
the density of the thermal component. In general one should not however  
ignore the fact that even at zero temperature the total density
 is different from the condensate density since the occupation numbers $n_i$ with $i\ne 0$
 of eq.(\ref{n1T}) differ from zero, giving rise to the quantum depletion of the
 condensate.  An example where quantum depletion plays a crucial role
 is presented in fig.3 which reports  the density profile of a cluster of 
 helium atoms calculated at $T=0$ using a correlated basis approach \cite{panda}. These systems, differently 
 from atomic gases, are strongly correlated. The figure clearly shows that
 the contribution of the
condensate is only a small fraction (about $10 \%$) and that 
in this case the measurement of the density profile 
would not yield any useful information on the condensate density $n_0$ even at
zero temperature. In the calculation of ref. \cite{panda} the condensate density
has been evaluated by determining the natural orbits which
diagonalize the 1-body density matrix (see eq.(\ref{n1T}))

\subsection{Equation for the order parameter}
  
The equation for the order parameter can be derived starting from the
Heisenberg equation for the field operator
\begin{eqnarray}
i\hbar \frac{\partial }{\partial t}\hat{\Psi}({\bf r},t) &=&[\hat{\Psi}({\bf %
r},t),\hat{H}]={\Big [}-\frac{\hbar ^{2}\nabla ^{2}}{2m}+V_{ext}({\bf r},t) 
\nonumber \\
&&{}+\int \hat{\Psi}^{\dagger }({\bf r}^{\prime },t)V({\bf r}^{\prime }-{\bf %
r})\hat{\Psi}({\bf r}^{\prime },t)d{\bf r}^{\prime }{\Big ]}\hat{\Psi}({\bf r%
},t)\;.  \label{H}
\end{eqnarray}

This is an exact equation if one uses for $V(s)$ the exact two-body 
interaction between particles and holds not only for dilute gases, 
but also for strongly interacting systems like superfluid $^4$He. A 
closed equation for the order parameter is obtained if we replace 
the field operator $\hat{\Psi}$ with 
the classical field $\Psi_0$ or, equivalently, with its expectation 
value (\ref{OP2}). This procedure is correct provided a series of 
assumptions are satisfied:

i) The thermal depletion should be small so that the thermal 
fluctuations of the field operator can be ignored. This implies 
temperatures much smaller than the critical value.

ii) The system should be so dilute  that the quantum fluctuations 
 of the field operator can be ignored. This requires that the 
gas parameter $na^3$  be much smaller than unity or, equivalently, that 
the average distance between particles  be much larger than the 
scattering length.

iii) The microscopic interaction $V(s)$ should be replaced by the 
pseudopotential $g4\pi\hbar^2(a/m)\delta({\bf s})$ fixed by the $s$-wave 
scattering length $a$. This implies that only low energy features of 
the problem can be investigated with this approach. Equivalently,
only variations of 
the order parameter over distances larger than the range of the 
interaction can be explored. 

iv) The number of atoms $N$ should be large enough in order to 
justify the Bogoliubov prescription. 

Under the hypothesis i)-iv)  one  obtains the
most famous Gross-Pitaevskii equation \cite{GP} 
\begin{equation}
i\hbar \frac{\partial }{\partial t}\Psi _{0}({\bf r},t)=\left( -\frac{\hbar
^{2}\nabla ^{2}}{2m}+V_{ext}({\bf r},t)+g\mid \Psi _{0}({\bf r},t)\mid
^{2}\right) \Psi _{0}({\bf r},t)\;,  \label{GP}
\end{equation}
for the classical field $\Psi_0$. This equation provides 
 the basic tool to explore the static and dynamic 
features of non uniform Bose gases at low temperature. The GP equation 
shares interesting analogies with  the Maxwell equations of classical 
electromagnetism. In both  approaches one
provides a description of a many-body problem (atoms anf photons respectively)
in terms of a classical field 
 (order parameter and electromagnetic field ). From this point of view a gas of  Bose-Einstein 
 condensed atoms can be regarded as a
{\bf classical matter wave}. With respect to the equations of  electromagnetism the 
GP equation however exhibits important differences: first it 
contains an important  non linear term due to interactions (this difference is not
crucial since non lineariteies
are exhibited also by electromagnetic  phenomena in dispersive media); 
 a second important difference is the role played by the Planck constant
  $\hbar$.  Differently from the Maxwell equations 
  the GP equation (\ref{GP})  in fact depends explicitly on this 
  fundamental constant of quantum mechanics.
  This feature is directly connected with the different 
   momentum-energy relationship exhibited in the two cases. 
   For  massive particles  the relation
   is quadratic ($E=p^2/2m$), while for photons it is linear ($E=cp$). When we employ the ondulatory description 
   through the de Broglie prescription  $E=\hbar \omega$ and 
   $p=\hbar k$, the relationship  bewteen 
   the wavevector and the frequency is still independent on the 
   Planck constant in the case case of photons ($\omega =c k$), 
   while it exhibits an explict dependence on $\hbar$ in the case 
   of atoms ($\omega=\hbar k^2/2m$). 
    This explains, for example, why interference 
    phenomena with classical matter waves depend explicitly on the 
    value of the Planck constant.

\subsection{Ground state configuration}

If we look for stationary solutions of the form
\begin{equation}
\Psi _{0}\left( {\bf r},t\right) =\Psi _{0}({\bf r})\exp (-i\mu t/\hbar )
\label{stationary}
\end{equation}
the Gross-Pitaevskii equation (\ref{GP}) 
becomes
\begin{equation}
\left( -\frac{\hbar ^{2}\nabla ^{2}}{2m}+V_{ext}({\bf r})-\mu +g\mid \Psi
_{0}({\bf r})\mid ^{2}\right) \Psi _{0}({\bf r})=0.  \label{GP1}
\end{equation}
This equation can be also derived  using the standard variational procedure
\begin{equation}
\delta(E-\mu N) =0
\label{varE}
\end{equation}
where 
\begin{equation}
E=\int \left( \frac{\hbar ^{2}}{2m}\mid \nabla \Psi_0 \mid ^{2}+V_{ext}({\bf %
r})\mid \Psi_0 \mid ^{2}+\frac{g}{2}\mid \Psi_0 \mid ^{4}\right) d{\bf r}
= E_{kin} + E_{ext} + E_{int} 
\label{E}
\end{equation}
is the energy functional of the system.
In this equation  we have naturally identify the kinetic energy ($E_{kin}$),
 the external potential  ($E_{ext}$) and the mean 
field interaction ($E_{int}$) energies.
Starting from the GP equation one can easily derive the useful relationship
\begin{equation}
N\mu = E_{kin} + E_{ext} + 2 E_{int}
\label{mu}
\end{equation}
for the chemical potential,  and 
the virial identity
 \begin{equation}
2E_{kin} - 2E_{ext} + 3E_{int} =0
\label{virial}
\end{equation}
characterizing the equilibrium configuration.
It is also
  useful to  recall  the so called release energy
\begin{equation}
E_{release} = E_{kin} + E_{int}
\label{release}
\end{equation}
 which coincides with 
  the energy of the system after switching off the trap and  is hence given
by the sum of the kinetic and interaction energies.

It is important to remind here that the quantity $E_{kin}$ 
entering the above equations is the kinetic energy of the condensate (mean field 
kinetic energy) and
should not be confused with the  full kinetic
 energy of the system. 
 The latter takes in fact contribution also from the particles out of the condensate and, using the notation
 (\ref{n1T}) for the 1-body density matrix, can
 be written in the form 
 \begin{equation}
 E_{kin}^{total} 
 = \int d{\bf r}{\hbar^2 \over 2m} \left( 
 \mid\nabla \Psi_0({\bf r})\mid^2   + \sum_{i\ne 0} 
n_i\mid \nabla \phi_i({\bf r})\mid^2\right) \, .
\label{Ekintotal}
\end{equation}
While for dilute gases the contribution $\sum_{i\ne 0}n_i$ 
of the non condensate component to the total number of atoms 
 is small at $T=0$ and one can safely identify $N$ with $N_0$,
the corresponding contribution to the kinetic energy (second term of 
eq.(\ref{Ekintotal})) may be much larger than the condensate kinetic energy. 
For similar reasons the mean field interaction energy $E_{int}$ should not be
confused with the full interaction energy whose evaluation requires the explicit knowledge of 
short range correlations.
  The  
 distinction between the {\it mean field} and  {\it total} energies becomes 
 particularly clear
 and stark in the case of a uniform gas interacting with hard sphere potentials. In this 
 case 
 the total kinetic energy coincides with the energy of the system, the interaction energy 
 being exactly
 zero due to the special choice of the  potential. Conversely 
 the kinetic energy of the condensate is zero because of the homogeneity of the sample. 
 Notice that
 the mean field scheme 
 fails to evaluate the full kinetic energy (\ref{Ekintotal}). This can be seen in the  case of
 a uniform gas where the occupation number of the single particle states
 is given by  the  Bogoliubov expression 
 $n_p = (p^2/2m +mc^2)/2\epsilon(p) -1/2$.
  The quantity $\sum_{p\ne 0}n_p$ is evaluable  
  and  gives the quantum depletion of the condensate. Viceversa
 the sum $\sum_{i\ne 0}(p^2/2m)n_p$ exhibits an ultraviolet divergency,
 showing that the evaluation of the kinetic energy requires  the proper knowledge of the
 momentum distribution at  momenta of the order of
 the inverse of the scattering length and cannot be achieved within the mean field 
 scheme. 
 
It is finally worth pointing out the key role played by the chemical potential  in 
the formalism of the GP equation. As one can see from (\ref{stationary}) the 
time dependence of the order parameter is fixed by
$\mu$ and not by the energy. This is a consequence of the fact that
 the order parameter is not a wave function and that the GP equation 
 is not a Schr\"odinger equation in the usual sense of quantum mechanics. 
 From the point of view of many-body theory the order parameter 
 corresponds to the matrix element of the field operator between two many-body 
 wave functions containing, respectively,  $N$ and $N+1$ particles. 
 This implies that its  time dependence is fixed by the factor $e^{-i(E(N+1) -E(N))t}$ 
 and hence by the chemical potential $\mu=\partial E/\partial N$ rather than by the energy $E$. 

If the Thomas Fermi parameter $Na/a_{ho}$ is very large the GP equation 
(\ref{GP}) can be solved analytically.  In fact in this case 
the density profile, as a consequence of the repulsive effect of the 
interactions, becomes so smooth that the kinetic energy term in (\ref{GP}) can be 
ignored and the density takes the shape of an inverted parabola (we consider
here spherical trapping)
\begin{equation}
n({\bf r}) = {1\over g} \left(\mu - V_{ext}\right) = 
{\mu\over g} \left(1 - {r^2 \over R^2}\right)
\label{nTF}
 \end{equation}
where the radius $R$  of the condensate is given by the formula
\begin{equation}
R = a_{ho}\left(15 N {a\over a_{ho}}\right)^{1/5}
\label{RTF}
\end{equation}
and $\mu = m\omega^2_{ho}R^2/2$. 
For an axially deformed trap the density profile is simply generalized to 
$n({\bf r}) =(\mu/ g) (1 - r^2_{\perp}/ R^2_{\perp}
- z^2 /Z^2)$
where $r^2_{\perp}=x^2+y^2$ and  the radial and axial radii are 
related by the equation $m\omega_{\perp}^2R^2_{\perp} = m\omega_z^2Z^2=2\mu$.
In many available configurations the size of the atomic cloud can  become much larger than the oscillator 
length and reach almost macroscopic values. 
 
\subsection{Visibility of the condensate}

In the introduction we have emphasized the important fact that, due to the
 harmonic trapping,   the condensate density can be separated from 
 the thermal component, a crucial condition for the experimental visibility
 of the condensate.  Due to the importance of this fact 
 it is worth 
 discussing in a deeper way the consequences of two-body 
 interactions. In fact the argument usually invoqued is based on the ideal gas model where
 the width of the
  condensate is fixed by the 
  oscillator length $a_{ho} = \sqrt{\hbar/m\omega_{ho}}$. On the other hand 
    the width of the thermal cloud is of the order of  
\begin{equation}
R_T \sim  \left( {k_B T \over m \omega_{ho}^2 }\right)^{1/2}
\label{RT}
\end{equation}
so that, for  "macroscopic" temperatures 
$ k_BT \gg \hbar \omega_{ho}$, one has $R_T \gg a_{ho}$ 
with the consequent net separability 
of the two components. 
 In the previous 
section we have however shown that interactions are responsible 
for a huge increase of the size of the condensate whose value, in the
 Thomas-Fermi limit, is given by eq.(\ref{RTF}). One then 
  conclude that interacations will partially reduce the 
 visibility of the condensate by making its size comparable to the
  one of the thermal cloud.  Interactions, by producing a smoother 
  density profile, make the system closer to a uniform gas and 
  consequently reduce the visibility of BEC in coordinate space. 
  In fig 4 we show a typical
  {\it in situ} measurement of the density profile taken at 
  different temperatures \cite{DW}.
  A similar argument holds if one explores the behaviour of the release 
energy $E_{release}$.
 For an ideal gas the release energy
is fixed by  the oscillator energy $\hbar \omega_0$,  which is much
smaller than  the release energy of the thermal cloud, fixed by $k_BT$.
 This provides a net separation between the two components which expand
 at  different velocities. The difference is sizably reduced if one 
 takes into account 
interactions which increase significantly the
 value of $E_{release}$ making it comparable 
 to the thermal energy $kT$. In conclusion  the  separability
of the condensate from the thermal component is significantly reduced by two-body interactions
and  requires a careful 
bimodal fit to the measured  profiles (both in {\it in situ} and after  expansion).

 While the visibility 
of the condensate in coordinate space is reduced by 
two-body (repulsive) interactions, it is  conversely enhanced  in momentum space. In fact, according
to the Heisenberg relationship, the width 
 of the condensate in momentum space behaves like 
$\hbar/R$ and hence becomes smaller and smaller as $R$ increases, 
approaching the typical $\delta$ distribution characterizing  
BEC in uniform systems.
While in the absence of interactions the visibility of BEC in coordinate  
and in momentum space are perfectly equivalent, due to the symmetric role 
played by the  spatial and momentum coordinates of  the harmonic Hamiltonian, in the 
Thomas-Fermi limit this symmetry is lost. The possibility of measuring 
the momentum distribution of trapped atomic clouds, via inelastic photon
 scattering experiments \cite{mitnp}  opens new possibilities to separate at a 
  deeper level the condensate from the thermal component.
In principle such measurements might provide an identification of the 
quantum depleted part too. In practice the effect of 
quantum depletion is too small to be observed in the presently available 
condensates. 

\subsection{Role of the phase: interference with BEC}

The  phase of the order parameter plays a crucial role in characterizing the interference
phenomena exhibited by a Bose-Einstein condensed gas.  The simplest example is the 
interference produced by  two initially separated condensates which expand 
and  overlap. The corresponding experiment was first carried out at Mit \cite{Mitinterf}. 

Let us first consider a single  expanding condensate. At large times the phase of the order parameter
takes a quadratic dependence, yielding the asymptotic form
\begin{equation}
\Psi_0({\bf r}, t) =  \mid \Psi_0({\bf r}, t)\mid e^{imr^2/2\hbar t}
\label{exp1}
\end{equation}
for the order parameter.
It is interesting to notice that the velocity field \begin{equation}
{\bf v} = {\hbar \over m} \nabla S 
\label{v}
\end{equation}
associated with the order parameter (\ref{exp1}) coincides with  the classical law
${\bf v} = {\bf r}/t$.  For large times the velocity then exhibits a 
fully classical behaviour. However the phase $S$ keeps its quantum nature fixed by the 
Planck constant.

If we now consider two expanding condensates
separated by a distance ${\bf d}$ and we assume that for large times the order parameter is given by 
a linear combination of the form
\begin{equation}
\Psi_0 =  \mid \Psi_1\mid e^{im({\bf  r}+{\bf d}/2)^2/(2\hbar t)} + e^{i\alpha} 
\mid \Psi_2\mid e^{im({\bf  r}-{\bf d}/2)^2/(2\hbar t)} 
\label{12}
\end{equation}
we find  that the density $\mid \Psi_0\mid^2$ acquires a modulation with typical 
fringes orthogonal to the vector ${\bf d}$ and separated by the distance
\begin{equation}
\lambda = {ht \over md} \, .
\label{lambda}
\end{equation}
This distance depends explictly on the Planck constant and with typical choices of  
the parameters $d$ and $t$ is a visibile length of the order of $10-20 \mu K$. 
The above estimate for the interference effects ignores the role of the interaction 
between the two clouds which are responsible, during the expansion, for deviations 
from the law (\ref{12}). The basic physics  is however accounted by this simple model which provides
a useful qualitative estimate of interference effects.
It points out, in particular,  a conceptually important question connected with the 
role of the relative  phase of the two condensates. Suppose that the two condensates are built 
in an independent way before  measurement. Why  should the order parameter
have the form (\ref{12}) with a well defined value of the relative phase $\alpha$ 
giving rise to interference? The question 
addresses the important problem of the quantum measurement
and of the consequent reduction of the wave packet. The standard theoretical  point view, based 
on the traditional rules of quantum mechanics, predicts that
after  measurement the wave function of the whole system will be of the form (\ref{12}) 
with a relative phase which cannot predicted in advance, unless the  condensates
were already  in a coherent configuration before 
the expansion.

\subsection{Role of the phase: irrotationality of the superfluid flow}

A  crucial consequence of the existence of an order parameter of the form (1),
is the irrotationality of the   flow. This feature is expressed by eq.(\ref{v}) from 
which one deduces that the phase plays the role of a velocity potential.
 Actually starting from the Gross-Pitaevskii equation for the order parameter it is
 possible to derive the equations of motion in a form which resembles the equations
 of irrotational hydrodynamics. These are the  basic 
 equations for the describing the motion of a superfluid. The derivation is 
 straightforward in a dilute gas. One starts from expression (1) for the order parameter and 
derives,  using  the GP equation (\ref{GP}), a closed set of equations 
for the density and for the phase or, equivalently, for the velocity field. The equations take the  form
\cite{ss}
\begin{equation}
\frac{\partial n}{\partial t}+{\rm div}({\bf v} n)=0\;,  \label{con}
\end{equation}

\begin{equation}
\hbar \frac{\partial }{\partial t}{S}+\left( \frac{1}{2}m{\bf v}%
^{2}+V_{ext}+gn-\frac{\hbar ^{2}}{2m\sqrt{n}}\nabla ^{2}\sqrt{n}\right)
=0.  \label{phi}
\end{equation}

The first equation is the usual equation of continuity, while the second one can be regarded as an 
equation for the velocity potential. Notice that in these equations, which are exactly equivalent to
the original Gross-Pitaevskii equation, the Planck
 constant enters through the so called quantum kinetic pressure term proportional to  $\nabla ^{2}\sqrt{n}$. 
 This term should be compared with the interaction 
  $gn$ given by two-body forces. If the role of interactions is very important the quantum pressure term 
 can be  neglected. This happens in the study of  the equilibrium 
  profile in  the Thomas-Fermi limit. The quantum pressure term is negligible also 
  in the time dependent problem, provided we study phenomena where the density varies 
  in space on lengths scales $L$ larger than the so called healing length
\begin{equation}
\xi = {\sqrt {\hbar \over 2mgn}} \, .
\label{xi}
\end{equation}
If $L \gg \xi$ it is immediate to see that the quantum pressure in (\ref{phi}) is 
negligible and that the same equation reduces to the simpler Euler-like form

\begin{equation}
m\frac{\partial }{\partial t}{{\bf v}}+\nabla \left( \frac{1}{2}m{\bf v}%
^{2}+V_{ext}+\mu \right) =0  \label{eu}
\end{equation}
where $\mu = gn$ is the chemical potential evaluated for a uniform body at the 
corresponding density.
The equation of continuity and the Euler equation (\ref{eu}) are called  
the hydrodynamic equations of superfluids. They are characterized by two important 
features:
the {\bf irrotationality} of the motion and  the 
crucial role of  {\bf interactions} which suppress the effects of the quantum pressure.  
It is worth noticing that the "hydrodyinamic" form of these equations is not the result of collisional
processes as happens in classical gases, but is the consequence of superfluidity.

It is also useful to recall that  the validity of the hydrodynamic equations is limited to
 the low $T$ regime (at finite temperature one should use the formalism of  two-fluid
 hydrodynamics). In the  limit of small temperatures these  equations can be used 
 to study also dense superfluids, like $^4$He as well as Fermi superfluids. Of course 
 in this case one should use the corresponding expression for the chemical potential. 
It is finally worth mentioning that in the hydrodynamic limit the Planck constant,
which entered equation (\ref{phi})  for the velocity field through the quantum pressure term,  
has disappeared from the equations of motion.

Starting from the hydrodynamic equations of superfluids one can explore different 
interesting problems in trapped BEC gases. These include the propagation of sound 
and the excitation of  collective modes. 

\subsection{Sound and collective excitations}

In the limit of small perturbations (linear limit) the 
hydrodynamic equations (\ref{con}) and (\ref{eu}) can be reduced to the typical  
form 
\begin{equation}
{\partial^2 \over \partial t^2 } \delta n = \nabla \cdot
[c^2({\bf r})\nabla \delta n]
\label{c2}
\end{equation} 
characterizing the propagation of sound waves in nonuniform media. Here $c({\bf r})$
is a ${\bf r}$-dependent 
sound velocity fixed by the relationship  $mc^2({\bf r}) = \mu - V_{ext}({\bf r})$. 
Sound can propagate in trapped Bose gases if the wavelength $\lambda$ of the 
wave is smaller than the size  of the system. In the
Thomas-Fermi limit this condition can be well satisfied especially in the elongated 
direction of the condensate. In the  experiments carried out at MIT \cite{Mitsound} 
the axial size $Z$  is a  few hundred 
micrometers while  $\xi$ is a fraction of micrometer, so  there is wide space for the 
propagation of sound waves in such samples. It is also interesting to consider the case where the
wave length $\lambda$ is smaller than the axial size ($\lambda < Z$), 
but larger than the radial size 
($\lambda > R_{\perp}$).
In this case the nature of the wave is characterized by 
 typical $1$-dimensional features \cite{Zaremba}.

If the wavelength becomes comparable to the size of the condensate the solutions of eq.(\ref{c2})
cannot be described in terms of a localized propagation of sound, but should be determined globally.
These solutions correspond to  oscillations of the whole system and the  discretization of the 
eigenvalues cannot be ignored in this case. 
For spherical trapping one can obtain analytic solutions of (\ref{c2})
in the form $\delta n = R_{n_r\ell}Y_{\ell m }$ where $\ell = 0,1,2..$ is the angular momentum (in units of $\hbar$) 
carried by
the excitation,  $n_r$ is the number of the radial nodes and $R_{n\ell}$
 is the radial function to be determined by solving the
equations of hydrodynamics. The corresponding eigenfrequencies obey the dispersion law 
\cite{ss}
\begin{equation}
\omega_{HD}= \omega_{ho} \sqrt{2n^2_r + 2n_r\ell + 3n_r + \ell}
\label{sandro}
\end{equation}
which should be compared with the prediction $\omega_{ho}(2n_r+\ell)$ 
of the ideal gas model. The role of
interactions is particularly evident for  the  surface excitations
($n_r=0$). In this case  
the hydrodynamic theory gives $\sqrt{\ell}\omega_{ho}$  to be compared with value $\ell \omega_0$ of
the ideal gas prediction.

Result (\ref{sandro}) reveals that, in the Thomas-Fermi 
limit $Na/a_{\rm ho}\gg 1$, the dispersion relation of the normal modes
of the condensate has changed significantly from the noninteracting 
behavior, as a consequence of two-body interactions. However it might 
appear surprising that in this limit the dispersion does not depend any 
more on the value of the interaction parameter $a$. This  differs from the 
uniform case where the dispersion law, in the corresponding phonon regime,
is given by $\omega =cq$ and depends explicitly on the interaction
through the  velocity of sound. The behavior exibited in  the harmonic
trap is well understood if  one notes that the discretized 
 values of $q$ are fixed by the
boundary and vary as $1/R$  where $R$ is the size of the system. While in
the box this size is fixed, in the case of harmonic confinement it increases
with $N$ due to the repulsive effect of two-body interactions (see eq.(\ref{RTF}): 
$R\sim (Na/a_{\rm ho})^{2/5}(m\omega_{\rm ho})^{-1/2}$. On the other hand the 
value of the sound velocity, calculated at the center of the trap, is 
given by  $c= (Na/a_{\rm ho})^{2/5}(\omega_{\rm ho}/m)^{1/2}$ and
also increases with $N$. One finally finds that  in the product $cq$ both
the interaction parameter and the number of  atoms  cancel out, 
so that the frequency 
  turns out to be  proportional to  the 
bare oscillator frequency $\omega_{\rm ho}$. 
A similar argument holds also for the surface excitations. In fact in the 
presence of an external force the dispersion relation of the surface modes obeys the
classical law $\omega^2 = F q/m$
where $F=m\omega_{ho}^2 R$ is the force evaluated at the surface of the system.
Since the product $\hbar qR$ gives the angular momentum carried by 
the surface wave, one immediately recovers the dispersion law $\sqrt{\ell}\omega_{ho}$.

Analytic solutions of the HD equation (\ref{c2}) are available also in the case of axi-symmetric
potentials of the form $V_{ext} = (m/2)(\omega^2_{\perp}r^2_{\perp} + \omega^2_z z^2)$. In this case
the third component  of angular momentum is still a good quantum number. 
For very elongated traps ($\omega_z \ll \omega_{\perp}$) 
the frequency of the lowest solution of even parity takes the value \cite{ss}
$\sqrt{5/2}\omega_z$ in excellent agreement with the experimental results
of \cite{Mitomega}.

The high precision of frequency  measurements is  not only providing us
with a powerful tool to check 
the predictions  of the hydrodynamic theory of superfluids, 
but  can be also used 
to explore finer effects, like the temperature dependence  
of the collective frequencies and, possibly,
beyond mean field corrections.

\subsection{Moment of inertia}

An important consequence of the hydrodynamic theory of superfluids is that the response 
\begin{equation}
\Theta = lim_{\Omega \to 0 }{<L_z> \over \Omega}
\label {Theta}
\end{equation}
to a rotating field of the form $H_{rot}=-\Omega L_z$, where $L_z= -i\hbar\sum_k{\bf r}_k\times{\bf \nabla}_k$ 
is the third component 
of the angular momentum operator, is given by the irrotational value
\begin{equation}
\Theta =\left({<x^2-y^2>\over <x^2+y^2>}\right)^2 \Theta_{rig}
\label {Thetairr}
\end{equation}
of the moment of inertia where $\Theta_{rig}$ is the classical rigid value. 
This result can be easily derived by rewriting
the equations of hydrodynamics in the frame rotating with the angular velocity $\Omega$.
Eq.(\ref{Thetairr}) can be also written in the form
$\Theta = \epsilon^2 \Theta_{irr}$ where $\epsilon =(\omega_x^2-\omega_y^2)/\omega_x^2+\omega_y^2)$
is the deformation of the trap in the plane of rotation
and we have used the Thomas-Fermi results for $<x^2>$ and $<y^2>$.

The quenching of the moment of inertia with respect to the rigid value was confirmed  in a series
of experiments of the 60's caried out on superfluid helium, providing   an 
independent measurement of the
superfluid density 
\cite{hess}. In this context it is worth mentioning that the superfluid density of superfluid
helium exhibits a very different behaviour with respect to the condensate density, especially at low 
temperature. In particular at $T=0$ the superfluid density coincides with the total density while the
condensate density is only  a small fraction ($\sim 10 \%$).

A challenging question is how to measure the moment of inertia of a trapped gas where the direct measurement 
of the angular momentum $L_z$ is not feasible due to the diluteness of the sample. An interesting possibility
is provided by the fact that, if the deformation of the trap is different from zero, 
the quadrupole and rotational degrees of freedom are coupled each other.
 This is well understood by considering
the exact commutation relation
\begin{equation}
[H,L_z] = im(\omega^2_x-\omega^2_y)Q
\label{comm}
\end{equation}
which explicitly points out the link between the angular momentum operator $L_z$ and the quadrupole
operator $Q=\sum_ix_iy_i$. 
Since the quadrupole variable can be easily excited and imaged
 one expects to obtain information on the 
rotational properties of the system by investigating
the quadrupole modes excited by the operator $Q$. An important case is the so 
called scissors mode. This
mode corresponds to the rotation of a deformed condensate whose 
inclination angle $\theta$ oscillates in time.
The oscillation can be induced by a sudden rotation of the confining trap with respect to the
initial equilibrium configuration. 
It is important to point out that the frequency of this oscillation does not vanish with $\epsilon$ as one would 
expect for a classical system. In fact in a superfluid both the restoring force parameter and the
mass parameter (moment of inertia) behave like $\epsilon^2$ so that the frequency keeps a constant value
when $\epsilon \to 0$. The hydrodynamic theory of superfluids predicts the 
value \cite{David} 
\begin{equation}
\omega_{HD}=\sqrt{\omega^2_x
+\omega_y^2 }
\label{scissors}
\end{equation}
in excellent agreement with the experimental results recently carried out at Oxford 
\cite{Foot} (see fig.5). 
Conversely in a classical gas one predicts
the value $\omega = \mid \omega_x\pm\omega_y\mid$. Also these values have been tested experimentally with high accuracy
by exciting the rotation of the gas at high temperatures, well above $T_c$. 

\subsection{Quantized vortices}

Quantized vortices are one of the most spectacular manifestations of superfluidity
(we will not discuss here other important manifestations of superfluidity, like the 
reduction of viscosity recently explored 
at MIT \cite{Mitviscosity}). The existence of quantized vortices is
the combined consequence of the behaviour of the phase of the order parameter,
which fixes the irrotationality
of the velocity field, and of the non-linearity of the equations 
of motion which is a crucial  the consequence of two-body 
interactions. In the Gross-Pitaevskii theory a quantized vortex 
can be regarded as a stationary solution of 
eq.(\ref{GP}) of the form
\begin{equation}
\Psi_0( {\bf r},t) = e^{-i\mu t}e^{i\phi}\Psi_v({\bf r})
\label{Phiv}
\end{equation}
where $\phi$ is the azimuthal angle and $\Psi_v$ is a real function obeying the equation
\begin{equation}
\left[ - {\frac{\hbar^2 \nabla^2 }{2m}} + {\frac{ \hbar^2 }{2m r_{\perp}^2}}
+ {\frac{m }{2}} (\omega_{\perp}^2 r_{\perp}^2 + \omega_z^2 z^2 ) + g
\Psi^2_v(r_\perp,z) \right] \Psi_v(r_\perp,z) = \mu \Psi_v(r_\perp,z) \;.
\label{GPV}
\end{equation}
The velocity field associated with the order parameter (\ref{Phiv}) 
takes the form ${\bf v} = (\hbar/m)\nabla \phi =
 \hat{\bf \Omega}\times {\bf r}/r^2$ where $\hat{\bf \Omega}$
 is the unit vector along
the $z$-th direction. 
It satisfies the irrotationality
constraint everywhere except along the
vortical line and gives rise to 
a total angular momentum given by 
\begin{equation}
L_z = N\hbar \, .
\label{Lz}
\end{equation}
The "centrifugal" term in $1/r^2_{\perp}$ of eq.(\ref{GPV}) originates from the 
peculiar behaviour of the velocity field of the vortical configuration.
This term is responsible
for the vanishing of the condensate density $\mid \Psi_v\mid^2$ along the 
$z$-th axis and produces the so called vortex core, whose size is fixed by the
healing length.
In fig.6 we plot a typical vortical profile obtained by solving the 
Gross-Pitaevskii equation (\ref{GPV}).

Quantized vortices are excited states of the system but they should not be confused with the elementary excitations
discussed in sect.I . From the many-body point of view the vortex correspond
to a state where all the N particles of the sample occupy the new solution of the GP equation. The excitation
energy and the angular momentum associated with the vortex 
are consequently "macroscopic" quantities while
the excitation energy and the 
angular momentum of elementary excitations are of the order, respectively, of the 
trapping frequencies  and of a few units of $\hbar$.

The excitation energy $E_v$ of the vortex, given by  
the difference between the vortex and 
the ground state energies, can be  calculated starting from 
the solution  of the
GP equation (\ref {GPV} ) and using expression (17) for the energy. For large
N samples the calculation  can be done analytically. One finds \cite{Lundh}
\begin{equation}
E_{v}=\frac{4\pi }{3}n_{0}Z{\frac{\hbar ^{2}}{m}}\ln {\frac{0.67R_{\perp }}{%
\xi }}
\label{Ev}
\end{equation}
where $n_0$ 
is the  value of the density in the center of the trap. 
It is worth noticing that the
factor in front of the logarithm  is proportional to
 the so called column density $\int dz n(x,y,z) = (4/3)Zn_0$ calculated along the 
symmetry axis ($x=y=0$).
As we will see later 
this identification is useful for the calculation of the the energy of
a vortex line displaced from the symmetry  axis. Result (\ref{Ev}) 
allows for  an estimate
of the critical angular velocity needed to generate an energetically stable vortex. 
In fact, calculating the energy
$H-\Omega L_z$ of the sample in the frame rotating with angular velocity ${\Omega}$,
 one easily finds that the vortex is the lowest energy
configuration if the angular velocity $\Omega$ exceeds the critical value 
\begin{equation}
\Omega_{cr} = {E_v\over L_z} = {\frac{5}{2}}{\frac{\hbar }{mR_{\perp }^{2}}}\ln {\frac{%
0.67R_{\perp }}{\xi }} \, .
\label{Omegacr}
\end{equation}
In deriving result (\ref{Omegacr}) we have used the  relationship 
$N=(8/15)\pi n_0R^2_{\perp}Z$.
Typical values of $\Omega_{cr}$ correspond to fractions of $\Omega_{\perp}$, the exact value depending on the geometry
of the trap and on the value of the Thomas-Fermi parameter. In the recent experiments carried out at ENS-Paris
\cite{Ensv}
with an enolangated  trap ($\omega_z/\omega_{\perp}\sim 1/10$) the calculated value 
of $\Omega_{cr}$ turns out to be about  
$\omega_{\perp}/3$. In this experiment the vortex is created by rotating a slightly asymmetric
trap at different values $\Omega$, in analogy with the rotating bucket experiment on superfluid 
helium. This procedure differes from the one followed at JILA \cite{jila2}
where the vortex was instead created with full optical methods. 
The value of the critical angular velocity
 measured at ENS is a factor 2 higher than the theoretical
 estimate (\ref{Omegacr}). The 
discrepancy is due to the fact that, even if the vortex is energetically favourable, its
nucleation is forbiddeen
by the occurrence of a barrier which disappears only at higher angular velocities. 
The nucleation of the vortex as
well as the study of its stability and life time is at present a topic of 
intense experimental and theoretical 
investigation.

An important question in the study of vortices is not only how to produce but also 
how to detect them.  The
size of the vortex core is in fact too small (fractions of microns) to be 
observed {\it in
situ}. It can be however observed
by imaging the atomic cloud after expansion. An alternative method is given by the 
measurement of angular momentum, 
a quantity which, according to eq.(\ref{Lz}), characterizes in a peculiar way the 
quantum nature of the vortex. The presence of angular momentum  produces a splitting of the 
quadrupole frequencies $\omega_{\pm}$ relative to the excitations with $\ell_z=\pm2$ (
$\ell_z$
is the third component of the agular momentum of the elementary excitation). 
The value of the splitting and its relation with the angular momentum $L_z$ of the
vortex
can  be
derived  using a sum rule approach. The result, in the Thomas-Fermi limit, 
is given by the formula
\cite{Zambelli}
\begin{equation}
\omega_+-\omega_-  = 2 {<L_z>\over m <r^2_{\perp}>} \, .
\label{omegapr}
\end{equation}
The splitting of the quadrupole frequencies $\omega_{\pm}$  produces 
a mechanism of precession  that can be detected
experimentally. To this purpose one generates a small quadrupole deformation in the plane orthogonal to
the vortex axis, by adding a field of the form $x^2-y^2$. The precession of 
the quadrupole deformation
is directly related to the splitting (\ref{omegapr}). In fact, 
by requiring that in the frame rotating with angular velocity $\Omega_{prec}$
the energies   $\omega_{\pm}$ of the two modes are degenerate, one finds 
the result 
$\Omega_{prec}=(\omega_+-\omega_-)/4$. Using the above 
procedure the authors of ref \cite{ens2} 
have observed a jump in the angular momentum of the condensate 
above a critical value of the   
stirring frequency (see fig. 7). This approach has permitted to test
the quantization of the angular momentum of a single vortex line.
 
Gyroscopic effects of different type have  
been observed in \cite{jila2} where the  precession of a vortex
line circulating around the symmetry axis of the condensate was investigated. 
The angular velocity characterizing this 
precession can be estimated by evaluating the energy and the angular momentum of 
a vortex line as a function of its distance $d$ from 
the symmetry axis. This calculation is easy 
if one  assumes that the vortex  line is straight. 
The $d$ dependence of the energy is determined, within 
logarithmic accuracy, by  the $d$-dependence of the column density 
$\int d{\bf r} n(x,y,z)$  characterizing the prefactor
of eq.(\ref{Ev}). This yields \cite{fetter}
\begin{equation}
E_v = N\hbar \Omega_c\left(1-{d^2\over R_{\perp}^2}\right )^{3/2}
\label{Evd}
\end{equation}
where $\Omega_c$ is the critical angular velocity (\ref{Omegacr}).
Using Stokes' theorem and working in the Thomas-Fermi limit 
it is possible to calculate also the angular momentum for which one finds the result
\begin{equation}
L_z= N\hbar \left(1-{d^2 \over R_{\perp}^2}\right )^{5/2} \, .
\label{Lzd}
\end{equation}
The angular velocity characterizing the precession is given by
$\Omega_{prec} = \partial E_v/\partial L_z$. For small values of the vortex displacement
($d \ll R_{\perp}$) one obtains the result \cite{Fetter} $\Omega_{prec}=(3/5)\Omega_{cr}$ 
which agrees reasonably well with the  experimental results of JILA \cite{Jilaprec}.

\subsection{Conclusions and acknowledgements}

The main purpose of these notes was to give an introduction to some key
 features which have been the object
of recent experimental and theoretical work in the physics of trapped 
Bose-Einstein condensates. Special emphasis 
was given to the interplay between the phase of the order
 parameter and two-body 
interactions
which characterize the non linearity of the Gross-Pitaevskii equation. This interplay is at the basis of
the superfluid behaviour exhibited by these systems and is responsible, in particular, for the existence of 
quantized vortices.

It is a pleasure to thank the fruitful  collaboration
  of the BEC  Trento  team and  the warm hospitality of the 
French friends  at the Institute d'Etudes Sientifiques de  Carg\`ese.

 \noindent
FIGURE CAPTION:

\noindent
{Fig.1.  Condensate fraction of a trapped Bose gas as a function of $T/T_c$. Circles are
the experimental results of \cite{Ensher}, while the dashed line is eq.(\ref{N0}).}

\noindent
{Fig.2. Condensate fraction as a function of temperature 
in  superfluid $^4$He (from \cite{Sokol}).}

\noindent
{Fig.3.  Total density and condensate density  profiles of a helium droplet of $70$
atoms (from \cite{panda}).}

\noindent
{Fig.4. Axial profiles of a cloud of sodium atoms at different temperatures (from
\cite{DW}).}

\noindent
{Fig.5. Scissors oscillation observed in \cite{Foot}. The full line corresponds to
a sinusoidal fit with $\omega/2\pi = 265 \pm 0.8 \, Hz$ to be compared with the theoretical 
prediction $265 \, Hz$ of eq.\ref{scissors}.}

\noindent
{Fig.6.  Density profile of a quantized vortex (from \cite{RMP}).}

\noindent
{Fig.7. Angular momentum per particle of a trapped condensate 
as a function of the angular velocity of the trap (from \cite{ens2}).}

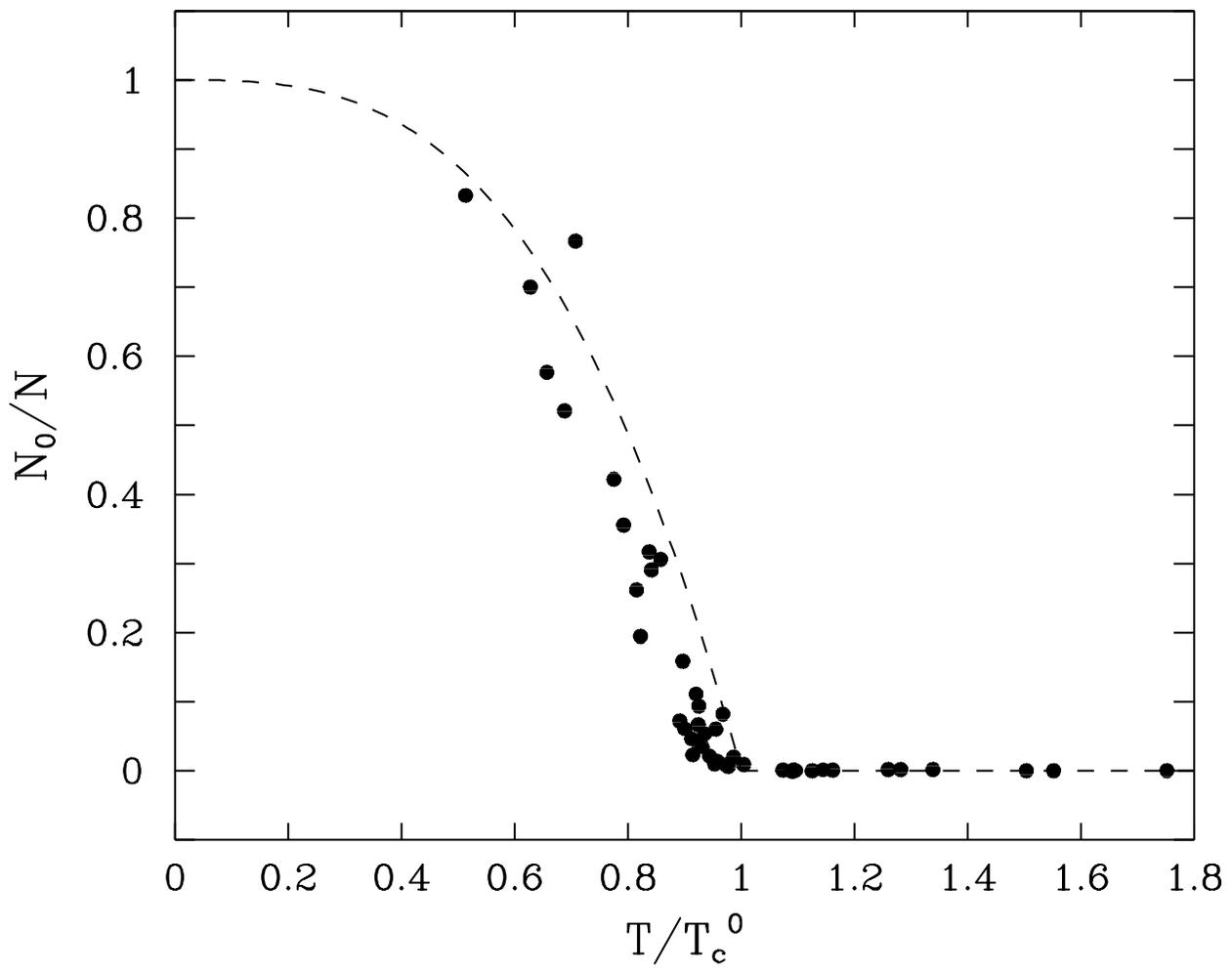
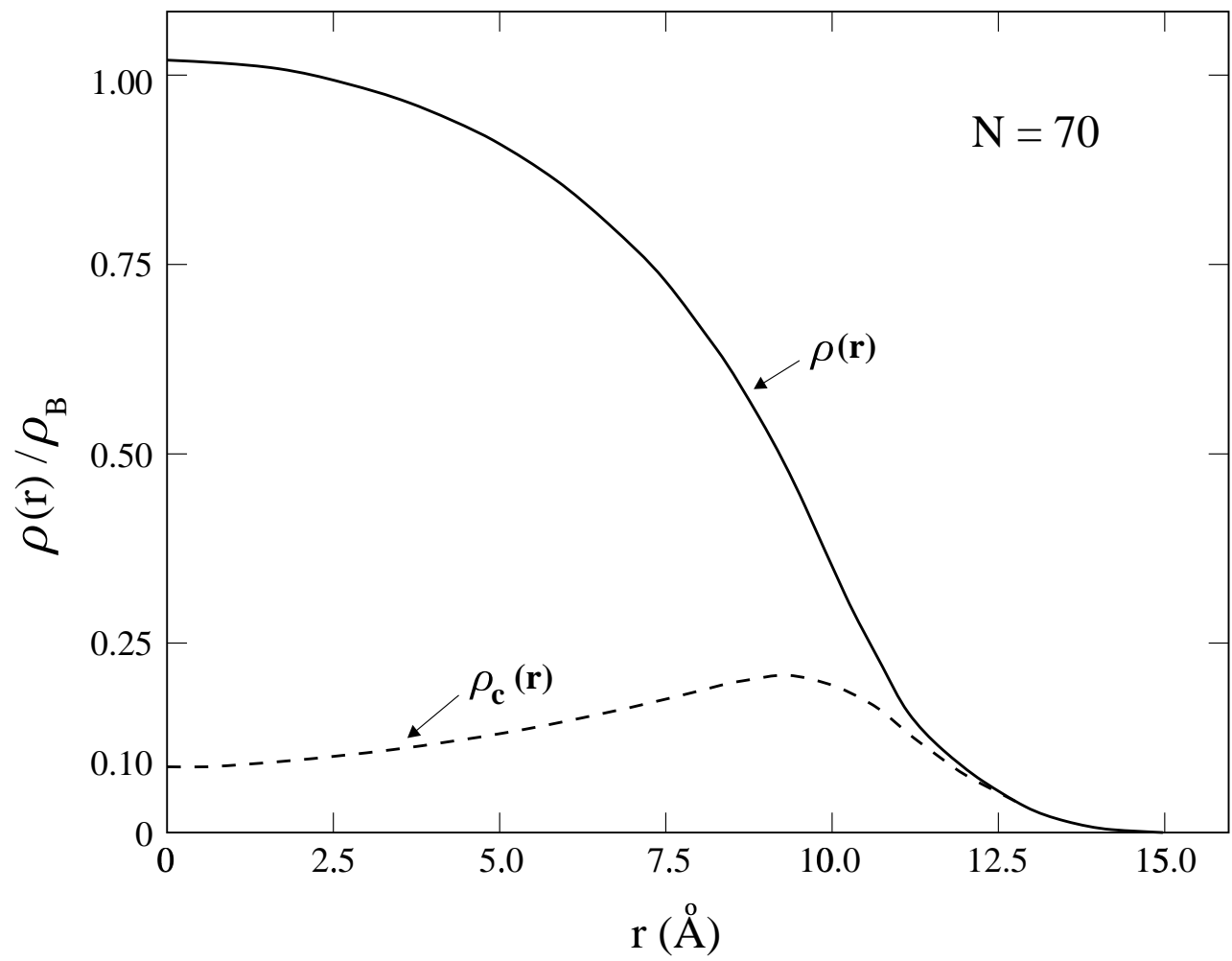
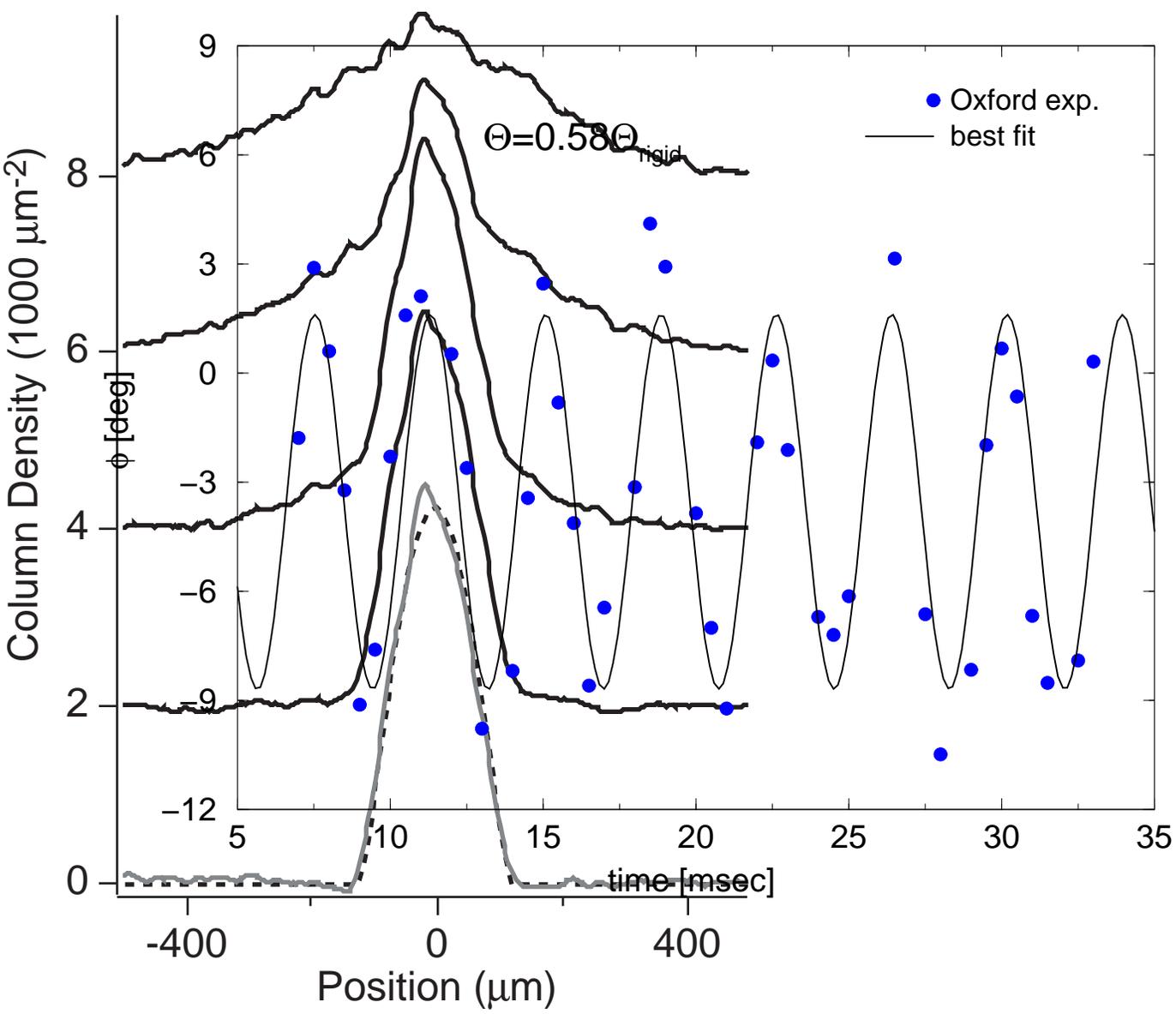
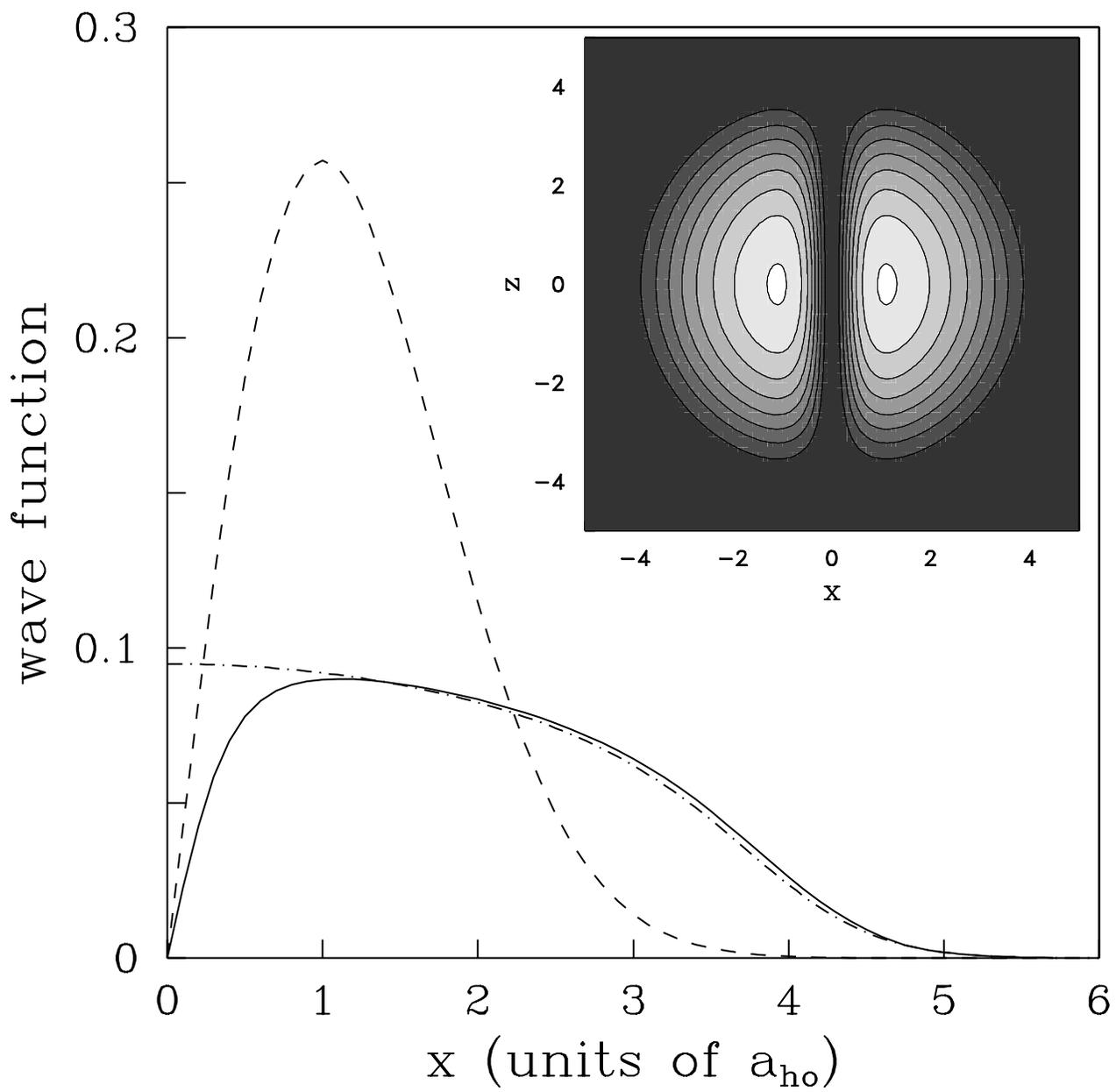
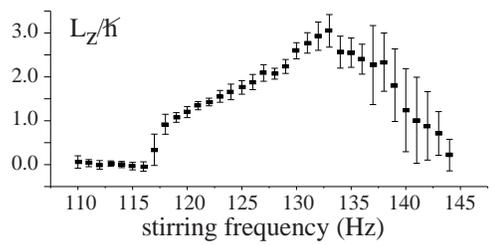
\end{document}